# Experiments with rotating collimators cutting out pencil of α-particles at radioactive decay of $^{239}$Pu evidence sharp anisotropy of space.

**S. E. Shnoll[1,2], I.A.Rubinshtejn[3], K. I. Zenchenko[2], V.A.Shlekhtarev[2], A.V.Kaminsky, A.A.Konradov[4], N.V.Udaltsova[2].**

[1]Lomonosov State University, Physical Department, Moscow, 119992, Russia

[2]Institute of Theoretical and Experimental Biophysics, Russian Academy of Sciences, Pushchino, Moscow Region, 142290, Russia

[3]Skobeltsyn Institute of Nuclear Physics, Moscow State University, Moscow, 119992, Russia

[4]Institute of Biochemical Physics, Russian Academy of Sciences, Moscow, Russia

**Summary**

As shown in our previous experiments fine structure of histograms of α-activity measurements serve as a sensitive tool for investigation of cosmo-physical influences. Particularly, the histograms structure is changed with the period equal to sidereal (1436 min) and solar (1440) day. It is similar with the high probability in different geographic points at the same local (longitude) time. More recently investigations were carried out with collimators, cutting out separate flows of total α-particles flying out at radioactive decay of $^{239}$Pu. These experiments revealed sharp dependence the histogram structure on the direction of α-particles flow.

In the presented work measurements were made with collimators rotating in the plane of sky equator. It was shown that during rotation the shape of histograms changes with periods determined by number of revolution. These results correspond to the assumption that the histogram shapes are determined by a picture of the celestial sphere, and also by interposition of the Earth, the Sun and the Moon.

## Introduction

It has been earlier shown, that the fine structure of statistical distributions of measurement results of processes of various nature depends on cosmo-physical factors. The shape of corresponding histograms changes with the period equal to sidereal and solar day, i.e. 1436 and 1440 minutes [1-4].

These periods disappeared at measurements of alpha-activity of $^{239}$**Pu** samples near the North Pole [5]. These results corresponded to the assumption of association of the histogram shapes with a picture of the celestial sphere, and also with interposition of the Earth, the Sun and the Moon.

However, at measurements at latitude 54°N (in Pushchino), absence of the daily period [6] also was revealed when using collimators restricting a flow of the alpha particles of radioactive decay at the direction to the north celestial pole. This result meant, that the question is not about dependence on a picture of the celestial sphere above a place of measurements, but about a direction of alpha particles flow.

In experiments with two collimators, directed one to the East and another to the West, it was revealed, that histograms of the similar shape at measurements with west collimator appear at 718 minutes (half of sidereal day) later then ones registered with East collimator [6]. Therefore, as acquired, the space surrounding the Earth is highly anisotropic, and this anisotropy is connected basically to a picture of the celestial sphere (sphere of distant stars).

This suggestion has been confirmed in experiments with collimators, rotated counter-clockwise, west to east (i.e. in a direction of rotation of the Earth), as well as clockwise (east to west). The description of these experiments is given further.

## Methods

As well as earlier, the basic object of these of research was a set of histograms constructed by results of measurements of alpha-activity of samples $^{239}$**Pu**.

Experimental methods, the devices for alpha-radioactivity measurements of $^{239}$**Pu** samples with collimators, and also construction of histograms and analysis of its shapes, are described in details in the earlier publications [2,3,8]. Measurements of number of events of radioactive decay were completed by device designed by one of the authors (I.A.R.). In this device the semi-conductor detector (photo diode) is placed after collimator, restricting a flow of the alpha particles in a certain direction. Results of measurements, consecutive numbers of events of the decay registered by the detector in 1-second intervals, are stored in computer archive.

Depending on specific targets, a time sequence of 1-second measurements was summarized to consecutive values of activity for 6, 15 or 60 seconds. Obtained time series were separated into consecutive pieces of 60 numbers in each. A histogram was built for each piece of 60 numbers. Histograms were smoothed using the method of moving averages for the greater convenience of a visual estimation of similarity of their shapes (more details see in [8,9]). Comparison of histograms was performed using auxiliary computer program by Edwin Pozharski [8].

A mechanical device designed by one of the authors (V.A.Sh) was used in experiments with rotation of collimators. In this device the measuring piece of equipment with collimator was attached to the platform rotated in a plane of Celestial Equator.

## Results

**A. Three revolutions of collimator counter-clockwise in a** day.

The diurnal period of increase in frequency of histograms with similar shape means dependence of an observable picture on rotation of the Earth.

The period of approximately 24 hours or with higher resolution 1436 minutes is also observed at measurements using collimators restricting a flow of alpha particles in a certain direction [6,7]. Therefore, the fine structure of distribution of results of measurements depends on what site of celestial sphere the flow of alpha particles is directed to. Studies of shapes of histograms constructed by results of measurements using rotated collimators testify to the benefit of this assumption.

The number of the "diurnal" cycles at clockwise rotation should be one less then numbers of collimator revolutions because of compensation of the Earth rotation.

At May 28 through June 10, 2004, we have performed measurements of alpha-activity of a sample 239Pu at 3 collimator revolutions a day, and also, for the control, simultaneous measurements with motionless collimator, directed to the West. Results of these measurements are presented on fig. 1-4. At these figures a dependence of frequency histograms of the same shape on size of time interval between similar histograms is shown.

Fig.1 shows results of comparison of 60-minute histograms, constructed at measurements with motionless collimator. A typical dependence repeatedly obtained in earlier studies is visible at the fig.1: histograms of the same shape most likely appear at the nearest intervals of time ("effect of a near zone") and in one day (24 hours).

Fig. 2 presents the result of comparison of 60-minute histograms constructed at measurements with collimator rotated 3 times a day counter-clockwise in a plane of celestial equator.

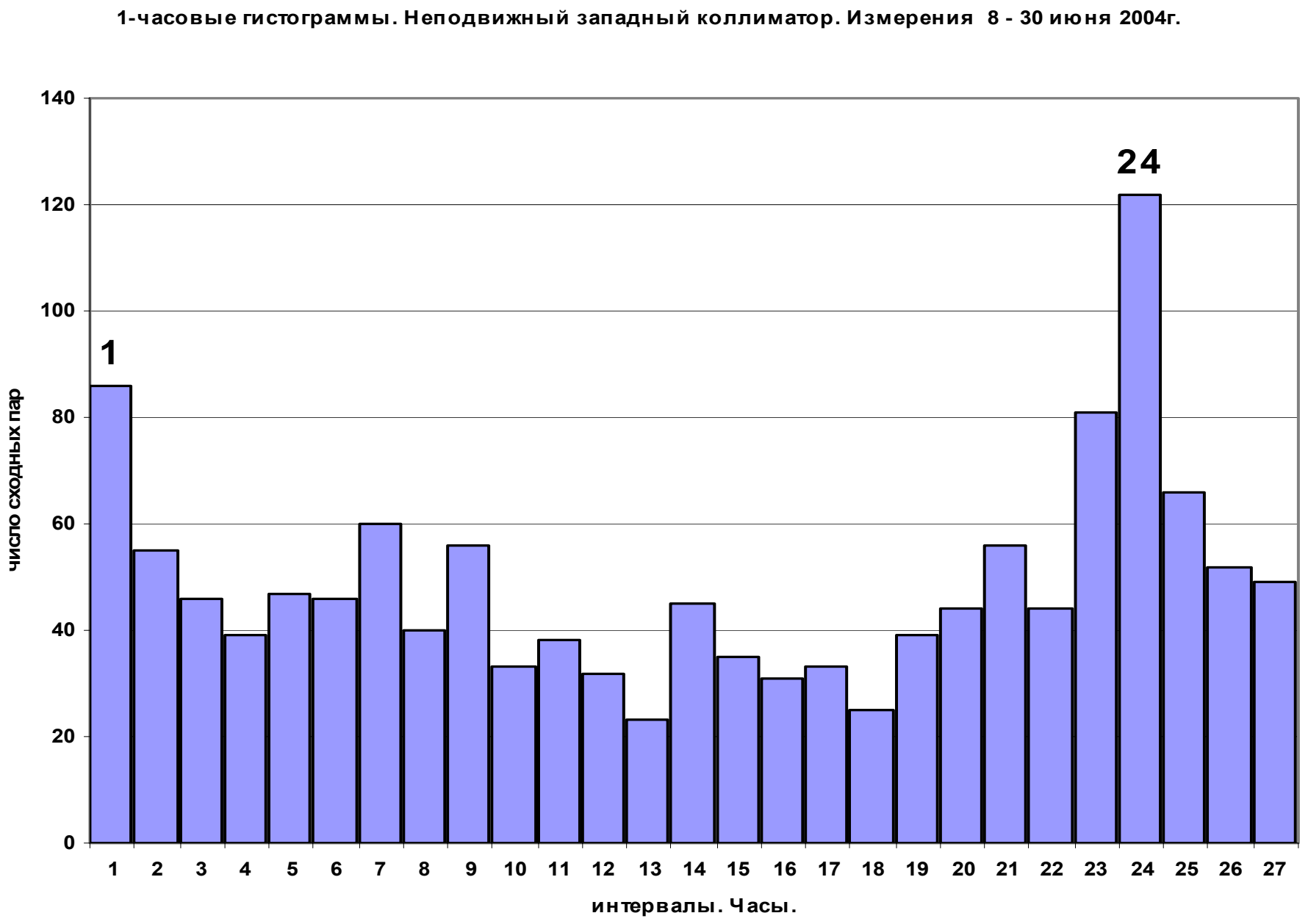

**Fig. 1. Frequency of similar 60-minute histograms against the time interval between histograms. Measurements of alpha-activity of a $^{239}$Pu sample by detector with motionless collimator directed to the West, June 8 – 30, 2004. Abscissa is time interval in hours. Ordinate is number of similar histogram pairs.**

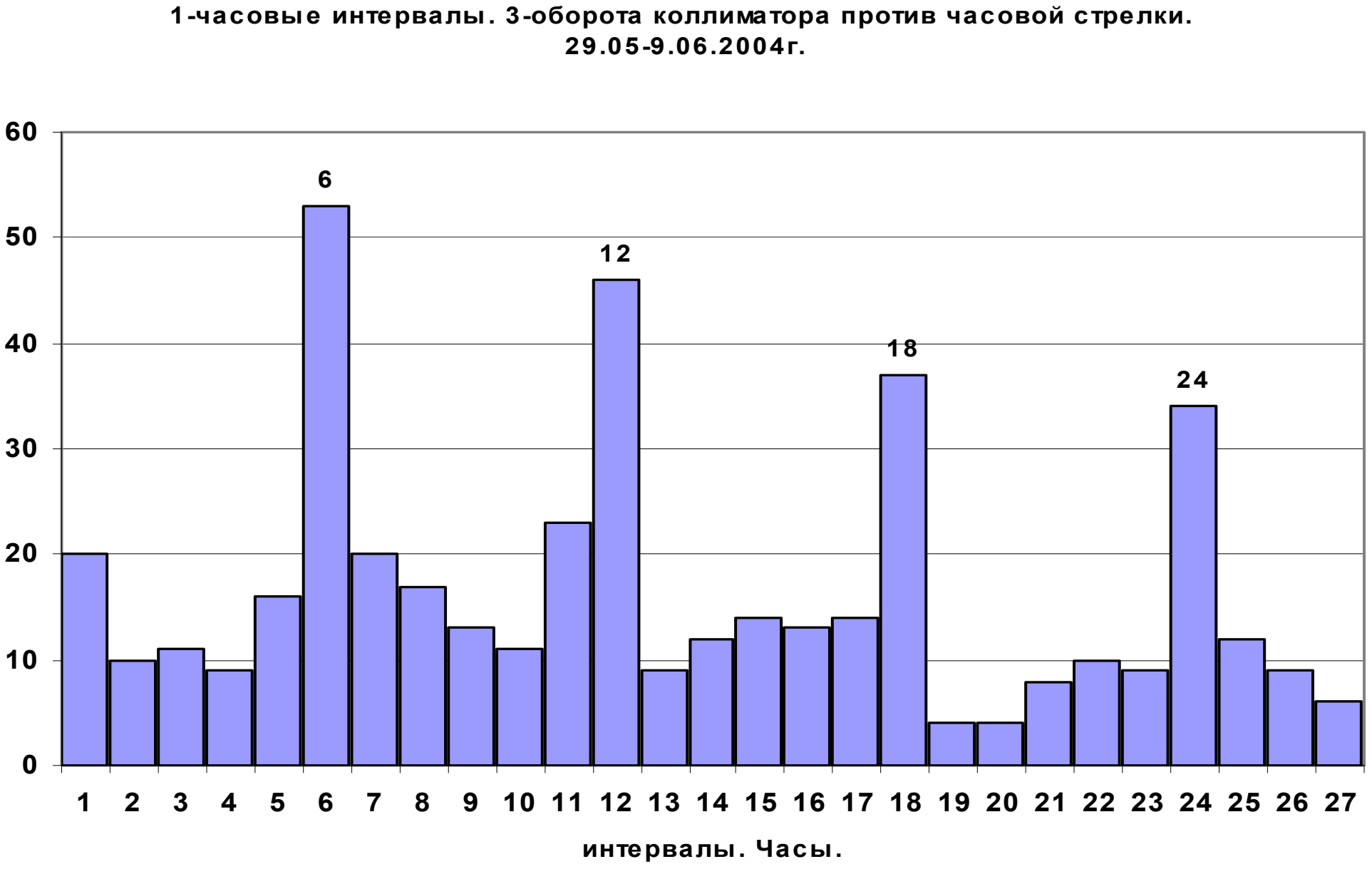

**Fig. 2. Frequency of similar 60-minute histograms against the time interval between histograms. Measurements of alpha-activity of a $^{239}$Pu sample by detector with collimator, making three revolutions counter-clockwise (west to east) in a day. Abscissa is time interval in hours. Ordinate is number of similar histogram pairs.**

As you can see at the fig. 2, at three revolutions of collimator counter-clockwise, the frequency of similar histograms fluctuates with the period of 6 hours: peaks correspond to the intervals of 6, 12, 18 and 24 hours.

24-hour period at a higher resolution consists of two components. It is visible by comparison of one-minute histograms shown at fig. 3 for measurements with motionless collimator and at fig. 4 for measurements at 3 collimator revolutions counter-clockwise. At measurements with motionless collimator (fig. 3) there are two peaks - one corresponds to sidereal day (1436 minutes), the second, which is less expressed, corresponds to solar day (1440 minutes).

You can see at fig.4 that 6-hour period at measurements with three revolutions of collimator also has two components. The first 6-hour maximum has two joint peaks of 359 and 360 minutes. The second 12-hour maximum has two peaks of 718 and 720 minutes. The third maximum (18 hours) has two peaks of 1077 and 1080 minutes. And the fourth one (24 hours) has two peaks of 1436 and 1440 minutes.

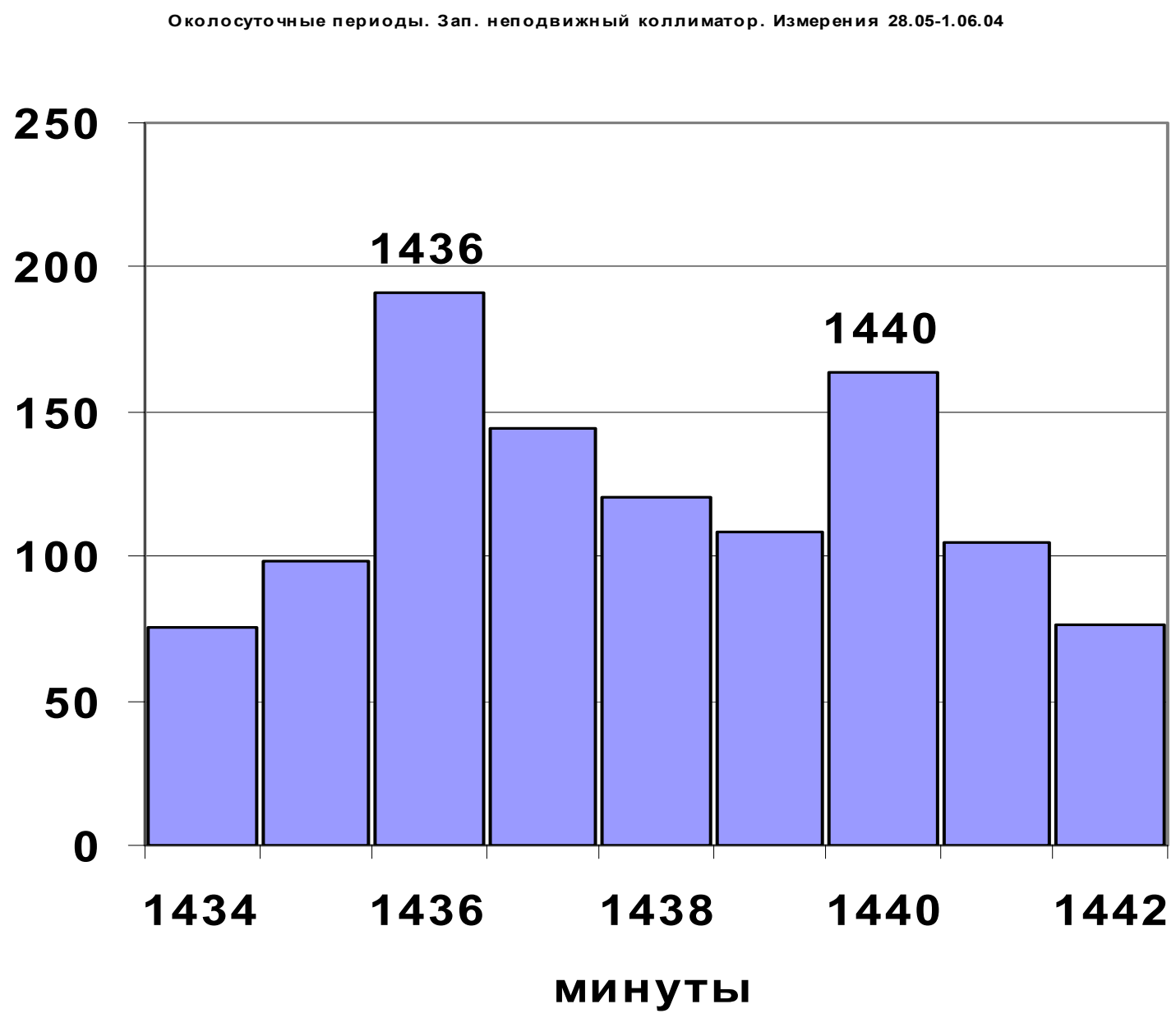

**Fig. 3. 24-hour period of frequency of similar histograms with the one-minute resolution. Measurements of May 29 - June 1, 2004 by detector with motionless collimator directed to the West. Abscissa is time interval in minutes. Ordinate is number of similar histogram pairs.**

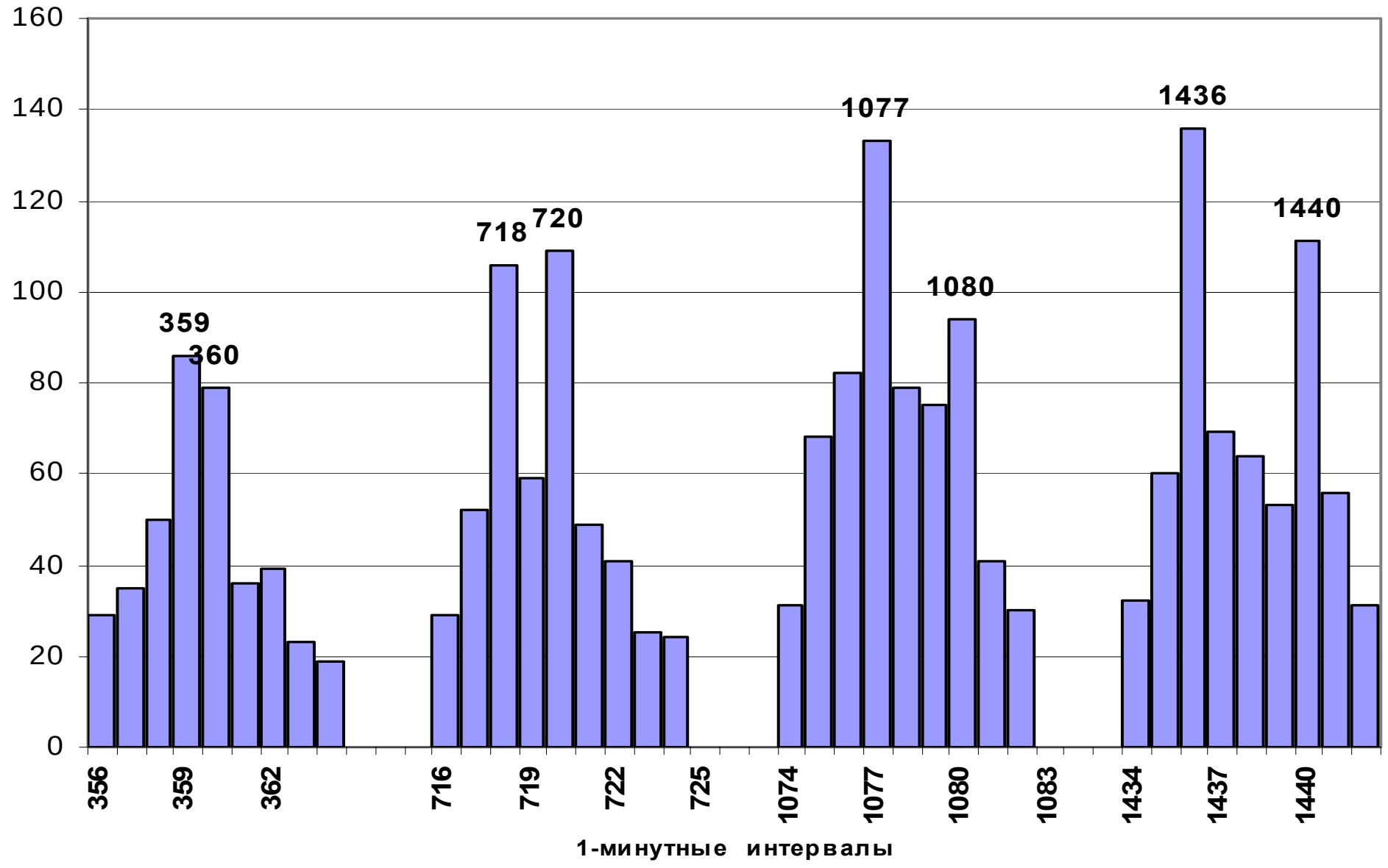

**Fig. 4. Experiments with rotated collimators. Frequency of similar 1-minute histograms by time interval between them. Three revolutions a day counter-clockwise. Two components of the 6-hour period: sidereal and solar.**

Results of these experiments confirm a conclusion according to which a change in histogram shape is caused by change in direction of alpha particles flow in relation to distant stars and the Sun (and other space objects). This conclusion is supported also by results of experiments with rotation of collimator clockwise.

In these experiments collimator made one revolution a day clockwise, east to west, i.e. against daily rotation of the Earth. As a result, the flow of alpha particles all the time was directed to the same point of celestial sphere. We expected in this case disappearance the diurnal period of frequency of similar histograms. This expectation was proved to be true.

On fig. 5 and 6 one can see that in such experiments frequency of appearance of similar 60 minute and 1-minute histograms doesn't depend on time. At the same time at synchronous measurements with motionless collimator the usual dependence with the diurnal period and near zone efect is observed.

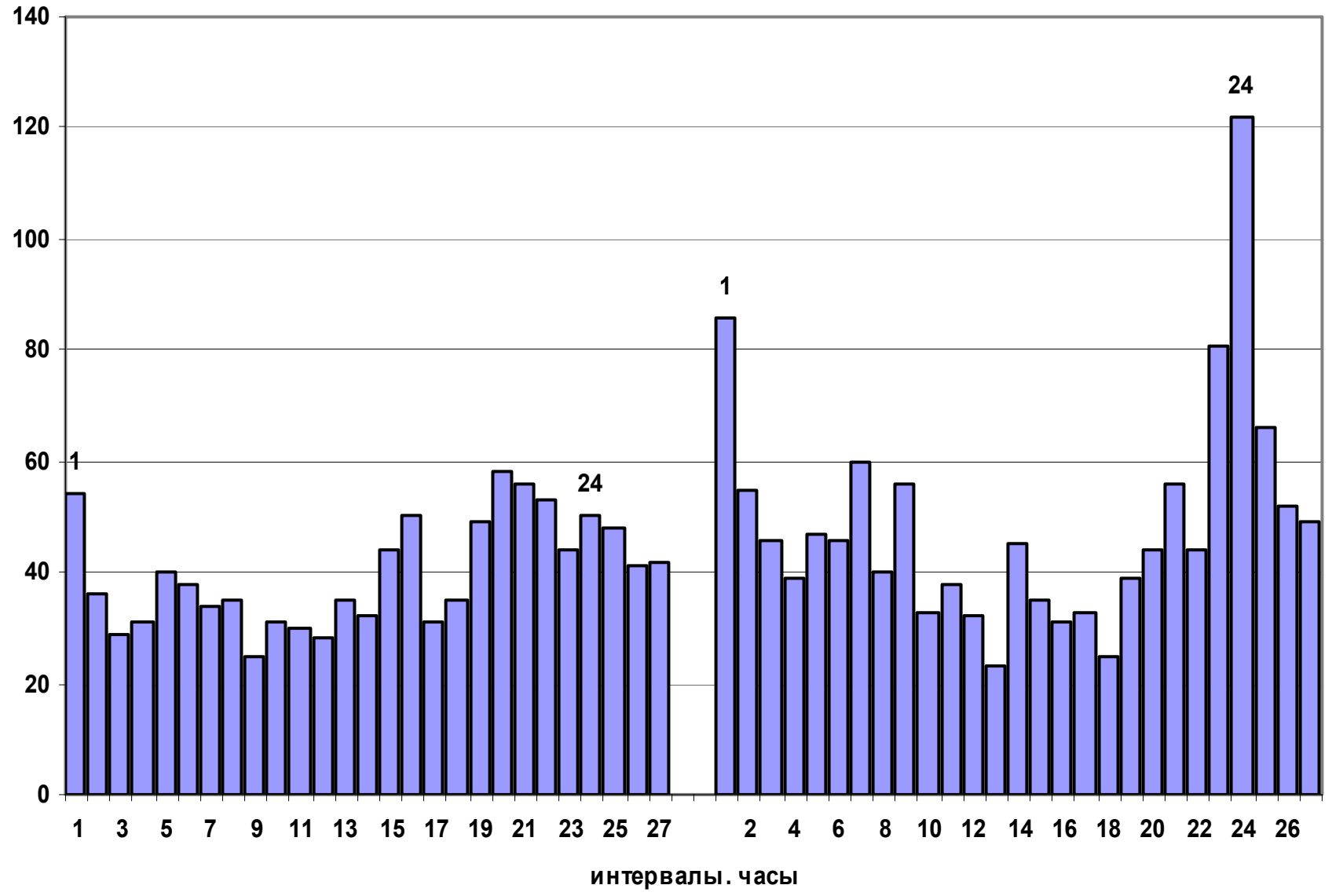

**Fig. 5. 60-minutes histograms. Left: 1 revolution clockwise. Right: control, motionless collimator**

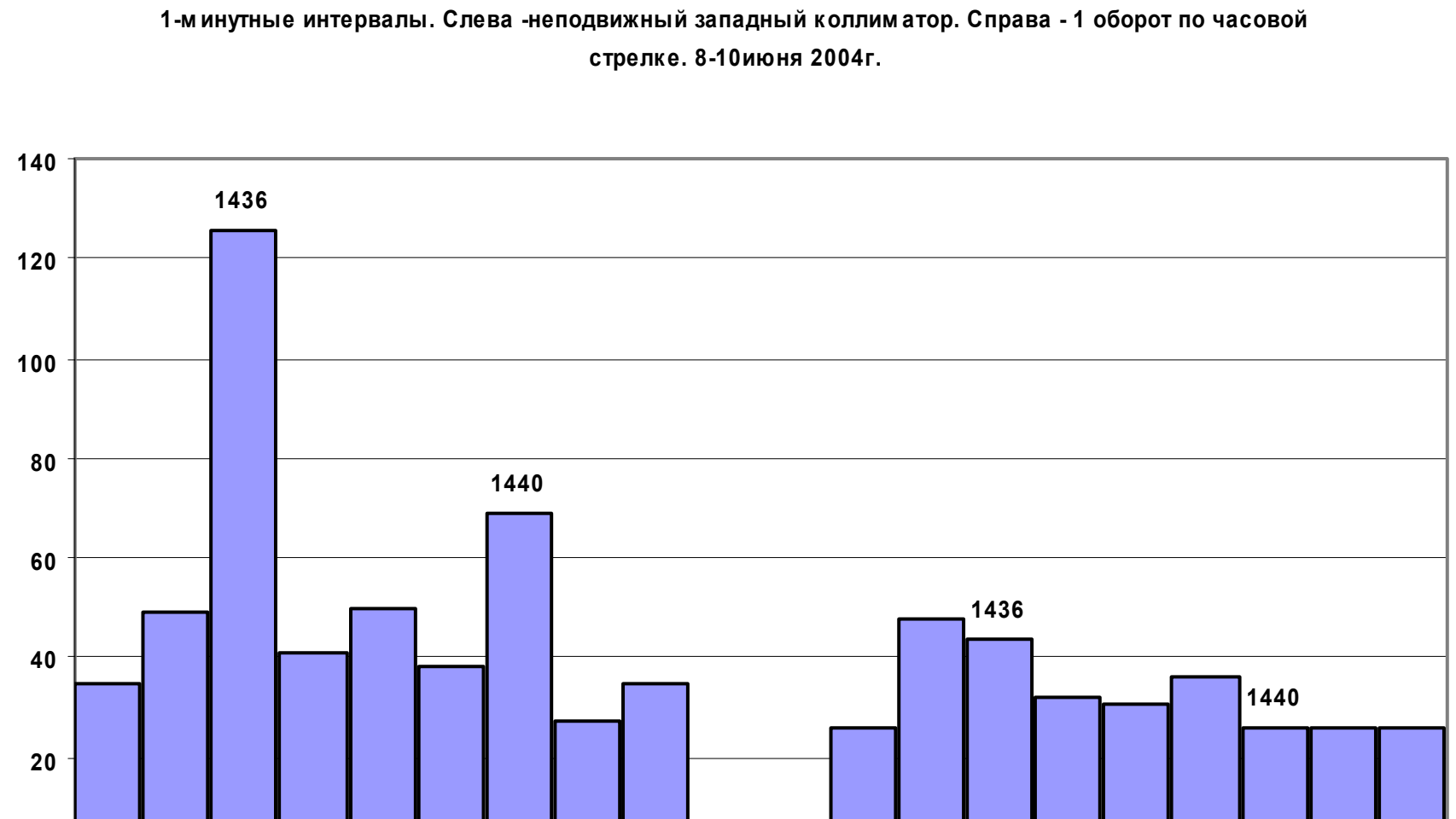

**Fig. 6. One-minute histograms. Left: control, motionless collimator. Right: rotation 1 revolution clockwise (east to west).**

## Discussion

Results of measurements with rotated collimator confirm a conclusion about dependence of fine structure of statistical distributions on a direction in space. This fine structure is defined by a spectrum of amplitudes of fluctuations of measured values. Presence of "peaks" and "hollows" at corresponding histograms suggests presence of the primary, allocated, "forbidden" and "permissible" values of amplitudes of fluctuations in each given moment [4]. Thus, a fine structure of statistical distributions presents a spectrum of the permissible amplitudes of fluctuations, and dependence of it on a direction in space shows sharp anisotropy of space.

It is necessary to emphasize, that the question is not about influence on the subject of measurement (in this case on radioactive decay). With accuracy of traditional statistical criteria, overall characteristics of distribution of radioactive decay measurements compliant with Poisson distribution [3]. Only the shape of histogram constructed for small sample size varies regularly. This regularity emerges in precise sidereal and solar periods of increase of frequency of similar histograms.

As shown above, the shape of histograms constructed by results of measurements of alpha-activity of samples **$^{239}$Pu,** varies with the period determined by number of revolutions in relation to celestial sphere and the Sun. In experiments with collimator, which made three revolutions counter-clockwise, the “diurnal” period was equal to 6 hours (three revolutions of collimator and one revolution of the Earth was observed - in total 4 revolutions in relation to celestial sphere and the Sun give the period equal 24/4 = 6 hours).

The result obtained in experiments with one revolution of collimator clockwise is not less important. The Earth rotation is compensated and a flow of alpha particles is directed all the time to the same point of celestial sphere. In these experiments the diurnal period was not observed at all.

The obtained results, though very clear ones, cause natural bewilderment.

Really, it is completely not obvious, by virtue of what reasons the spectrum of amplitudes of fluctuations of number of alpha particles, may depend on a direction of their flow in relation to celestial sphere and the Sun. The explanation of these phenomena probably demands essential change in general physical conceptions.

In such situation a dominant problem is to validate a reliability of the discussed phenomena. In aggregate of performed studies, we believe this task was completed .

ACKNOWLEDGEMENTS

S.E.Shnoll is indebted to M.N. Kondrashova and L.A. Blumenfeld for mutual understanding.

The authors are grateful to our colleagues, T.A. Zenchenko, D.P. Kharakoz B.M. Vladimirsky, B.V., Komberg, V. K. Lyapidevskii for collaboration and for valuable discussions.

The vivid interest of V.P.Tikhonov to the problem studied and his generous financial support were essential.

Correspondence and requests for materials should be addressed to S.E.S. (e-mail: shnoll@iteb.ru).